\documentclass[10pt,conference]{IEEEtran}
\usepackage{epsfig}
\usepackage{color,booktabs}
\usepackage{graphicx}
\usepackage[figuresright]{rotating}
\usepackage{amssymb}
\usepackage{amsmath}
\usepackage{setspace}
\usepackage{rotating}
\usepackage{multirow}
\usepackage{wrapfig}
\usepackage{booktabs}
\usepackage{array}
\usepackage{comment}
\usepackage{pict2e}
\usepackage{epstopdf}
\usepackage[ruled]{algorithm2e}

\usepackage{balance}
\usepackage{mathtools}
\usepackage{subcaption}
\usepackage[font=footnotesize]{caption}
\usepackage{xparse}

\usepackage{colortbl}

\NewDocumentCommand\jj{+u{\jj}}{\ignorespaces}

\begin{document}
\SetEndCharOfAlgoLine{}
\title{CollabLoc: Privacy-Preserving Multi-Modal Localization via Collaborative Information Fusion
}




\author{{Vidyasagar Sadhu, Dario Pompili, Saman Zonouz, Vincent Sritapan$^*$}\\
Department of Electrical and Computer Engineering, $^*$Cyber Security Division\\
Rutgers University, $^*$Department of Homeland Security Science \& Technology Directorate\\
\textit{ \{vidyasagar.sadhu, pompili, saman.zonouz\}@rutgers.edu}, \textit{vincent.sritapan@hq.dhs.gov}
}

\maketitle
\thispagestyle{empty}

\begin{abstract}
Mobile phones provide an excellent opportunity for building context-aware applications. 
In particular, location-based services are important context-aware services that are more and more used for enforcing security policies, for supporting indoor room navigation, and for providing personalized assistance. However, a major problem still remains unaddressed---the lack of solutions that work across buildings while not using additional infrastructure and also accounting for privacy and reliability needs. 
In this paper, a privacy-preserving, multi-modal, cross-building, collaborative localization platform is proposed based on Wi-Fi RSSI (existing infrastructure), Cellular RSSI, sound and light levels, that enables room-level localization as main application (though sub room level granularity is possible). The privacy is inherently built into the solution based on
onion routing, and perturbation/randomization techniques, and exploits the idea of weighted collaboration to increase the reliability as well as to limit the effect of noisy devices (due to sensor noise/privacy). The proposed solution has been analyzed in terms of privacy, accuracy, optimum parameters, and other overheads on location data collected at multiple indoor and outdoor locations. 
\end{abstract}



\section{Introduction}\label{sec:intro}
Mobile phones have become ubiquitous in our everyday lives. 
Today's research is progressing towards maximizing the potential benefits offered by mobile devices~\cite{viswanathan2012autonomic}. They range from outdoor navigation and real-time traffic prediction to remote patient monitoring and personal assistant technologies like Apple Siri and Google Now. 
\textbf{Motivation: }Many of today's location sensing techniques for location-based services still face many open-research challenges: 
%
1) \textit{Infrastructure:} Indoor localization solutions are not ubiquitous as of now due to the need for extra infrastructure or other complex requirements. 2) \textit{Very few cross-building solutions:} Majority of the localization solutions fall into either outdoor or indoor categories. While the former works only outdoor and the latter only indoor (works only for one building at a time), there are not many hybrid solutions that can work across buildings (possibly with reduced granularity such as room or building level). 3) \textit{Room-level granularity is sufficient:} In certain security applications such as~\cite{swirls}, policies are defined based on the device's context (location, time, etc.). Policies define allowed behavior such as whether recording or making a call is allowed or not and can depend on room-level location. To give an example, if a device is found to be in the ``Meeting Room,'' policy~A---which does not allow any calling or recording---will be enforced. Conversely, if a device is found to be in the ``Lobby,'' policy~B--- which does not impose any restriction on calling/recording---will be enforced. It is apparent how, in such applications, it is not necessary to have fine-grained accuracy as it is in indoor localization scenarios, and room-level granularity is sufficient. Secondly, we believe finding the correct room is the major task in indoor localization. Once the the correct room is identified, in most cases humans can themselves navigate the room without any external assistance. Unless needed, having fine-grained accuracy will come at an additional cost (e.g., power consumption). 4) \textit{Privacy:} Many of the existing localization solutions ignore this important aspect raising concerns among users. Users are worried that the locations obtained by these solutions will be used in ways that may compromise their privacy. For example, some people opt out of Google's Wi-Fi database~\cite{googlewifi} feature as they fear the location data gathered by Google can be used to localize them. 5) \textit{Reliability:} Location obtained through collaboration is likely to be more reliable than the one obtained by a single device---reasons being lack of enough sensors in a single device, sensor noise, etc. 

\begin{figure}
\begin{center}
\includegraphics[width=3.4in]{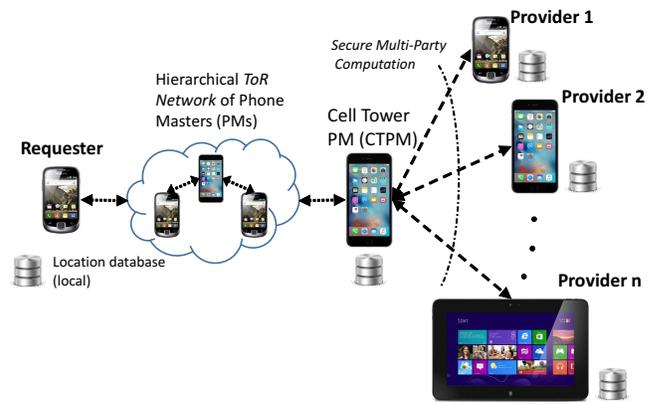}
\end{center}
\caption{An overview of CollabLoc framework.}\label{fig:collabloc}
\vspace{-0.2in}
\end{figure}

\textbf{Our Approach: } 
In order to address these challenges, we propose a unified multi-modal location determination framework (as a mobile application) that tries to address each of the above mentioned issues (Fig.~\ref{fig:collabloc}). Our framework can be readily deployed in the real world without the need for extra infrastructure (other than the already existing Wi-Fi Access Points~(APs)) as we base it on users' smartphones. Our solution is based on Wi-Fi Received Signal Strength Indicator~(RSSI) measurements but we augment it with additional sensor data such as sound level, light level, cell signal level, etc. to boost the granularity. This also helps in areas with lower Wi-Fi AP density and to distinguish among different regions/rooms within the area covered by a single (or set of) Wi-Fi AP(s). \textit{Though we target our solution mainly for room-level granularity}, the granularity can also be adjusted to be either more or less than room-level granularity by a parameter called Wi-Fi similarity threshold (see Sect.~\ref{sec:prop_soln}). As mentioned above, having more than needed granularity comes at an additional cost. As such, \textit{our solution neither competes nor is comparable with other indoor localization solutions}. Unlike indoor localization solutions, our solution is pervasive (i.e., not limited to a single building)---we do not build a map of the building or identify important fixtures (such as elevators). We tag different locations \textit{in and around} buildings by a location label using the location-specific signatures/features (such as Wi-Fi APs and their strengths, sound, light, cell signal levels, etc.) as tags. 
In addition, we make privacy as an integral part of our solution (rather than an addon) for which, we propose data perturbation and randomization techniques and also a ToR network~\cite{tor} composed of smartphones using our solution. To address the last issue on reliability, we obtain location through collaboration from multiple devices. Collaboration not only helps in improving accuracy (by reducing the amount of noise added by virtue of privacy) but also in knowledge sharing from one device to another. 
It is worth emphasizing that we have not used any local collaboration (i.e., collaboration among devices in the vicinity) as such approach will violate their privacy. Also, it is not necessary that all the people who know about that location are present in the vicinity. 
Additionally, we make use of weighted fusion to further improve accuracy.
To reduce energy footprint, we make use of Wi-Fi scan data most of the time, which is \textit{automatically} generated by the mobile Operating System~(OS) when the Wi-Fi is ON.
%


\textbf{Our Contributions: }
\begin{itemize}
\item We propose a \textit{privacy-preserving} multi-modal \textit{cross-building} room-level localization framework (sub-room level possible). To the author's best knowledge, ours is the first such solution. The various multi-modal features considered are Wi-Fi RSSI values, Cell ID, Location Area Code~(LAC), sound level and cell-signal RSSI values. Privacy is preserved via ToR routing and perturbation/randomization techniques. 
  \item We developed a custom two-step classifier, 
  Number of Feature Matches~(NFM) 
  that is shown to perform better than Multinomial Logistic Regression~(MLR). 
  \item We use the idea of weighted collaboration in order to increase accuracy as well as to filter out the noise introduced because of privacy requirements.  
  \item We evaluated our solution on real location data to show the trade-offs among accuracy, privacy and other overheads by developing an Android application to collect the above location features.
\end{itemize}

\textbf{Paper Outline: }The reminder of this paper is structured as follows. In Sect.~\ref{sec:rel_work} we compare our approach with some of the related work in location sensing, especially in collaborative sensing and privacy-preserving localization. In Sect.~\ref{sec:prop_soln}, we elaborate on different components of the proposed solution. 
In Sect.~\ref{sec:perf_eval}, we evaluate our solution both theoretically and experimentally on real location data.
Finally, in Sect.~\ref{sec:conc}, we conclude the paper 
and discuss future directions.

\section{Related Work}\label{sec:rel_work}
There are existing solutions that store databases of cell IDs and Wi-Fi Access points. An example of the former is the Open Cell ID project~\cite{opencellid}, which provides GPS positions of GSM cell stations based on Cell ID, Mobile Country Codes~(MCC), and Mobile Network Codes~(MNC). Examples of the latter are WiGLE~\cite{wiglewifidatabase} and Google Maps Geolocation API~\cite{googlewifi}, which provide the GPS coordinates (along with some radius of uncertainty) of the Wi-Fi access point from its SSID and MAC address. \jj Both these databases are formed from publicly sourced data.\jj Both these solutions are centralized and thereby prone to single point of failure and to scaling problems. In our solution the location data is distributed across mobile devices and is not stored at a single location. Also, in the case of Open Cell ID, the location is not known exactly but only to a cell tower level; whereas in the case of WiGLE, the major problem is privacy, because of which some people may not be willing to share their Wi-Fi AP details with others.


There are some privacy-preserving localization solutions in the literature. For example,
Gedik and Liu~\cite{kanonyloc} present a location-privacy method that makes use of general k-anonymity model.
Here, a person's location is indistinguishable from that of $k-1$ anonymous people around him/her. However, in our collaborative setting this would result in a large communication overhead as the locations of $k-1$ people have to be known. Kassem and Kang~\cite{Fawaz2014LocationUsers} provide techniques to address location tracking, profiling, and identification threats on Android OS. 
Conversely, since in our scenario more than one device is involved, we propose techniques to preserve privacy during collaboration between location providers and requester.
There is some existing work on local collaboration such as~\cite{collabcxtrecog2003}, \cite{darwinphones2010} to increase accuracy. These solutions neither preserve privacy nor consider the effect of noisy devices in the collaboration process. Also since collaboration is limited to only local devices, these approaches lack the advantage of our approach where
any device can become eligible for collaboration, provided it has some information about that place, or near by places. 

There is also existing literature on indoor localization and floor-map reconstruction with minimal infrastructure and human intervention.
For example, Wang et al.~\cite{Wang2012NoWar-drive} propose a solution using the combination of dead-reckoning, user-activity recognition from mobile sensors, and WiFi-based partitioning of an area. Rai et al.~\cite{Rai2012Zee:Localization} make use of crowdsourcing to gather Wi-Fi signatures and determine the signature location using sensor activity recognition and a map of the floor plan. These techniques, unlike ours, do not consider privacy, and are mostly limited to a single building (albeit achieving high granularity) and cannot scale to work across buildings.
There are also certain room-level localization solutions. For example, Shen et al.~\cite{Shen2016Feature-BasedSmartphones} propose a technique that uses a combination of RSSI measurements and room specific user activity and dwell times. Kyritsis et al.~\cite{Kyritsis2016AMethod} make use of RSSI readings from BLE beacons fixed in the rooms along with the geometry of the room. 
All of these solutions are designed to work only in a single or at most a few buildings and incur considerable overhead to pervasively work across buildings. Moreover they do not preserve user privacy. \textit{In summary, the following features distinguish our work from others---ability to work across multiple buildings (pervasive), privacy preserving and using collaboration to increase accuracy.} 


\section{Proposed Solution}\label{sec:prop_soln}
Our solution, which is available as a mobile application, can be divided into three components (see Fig.~\ref{fig:collabloc}). The first one involves each device building local database  
consisting of multi-modal location features and location labels; \textit{the labels are predominantly acquired from collaboration sessions}. The second component involves the location requester contacting the location providers in a secure and privacy-preserving manner via ToR network. The third component consists of fusing the results of multiple providers to return the result to the requester. Each of these components are explained below.

\textbf{Local Learning: }
Each mobile phone using our solution maintains a local database consisting of location features and the corresponding location labels. \textit{Local learning} is a continuous phase (happens continuously) where the device updates its location database with \textit{new} location features and labels. Location features consist of ``list of Wi-Fi Access Points~(APs),'' ``sound-level,'' ``light-level,'', ``geo-magnetic signal'', 
``cell tower ID,'' ``Location Area Code~(LAC),'' ``cell signal strength,'' etc. Examples of location labels can be ``Conference Room A" (in a conference hotel),'' ``Room 213" (in a library),'' etc. These location labels are publicly known names (i.e., not personalized ones like home, office, etc.) and entered manually \textit{only once} by the location natives.
To ensure uniformity among all labels and to remove ambiguity, we will make use of an ontology based framework~\cite{Wang2004OntologyOWL} for the labels that takes user input and converts them into standard labels that are hierarchically defined (e.g., country, zip code, street, building number, room name). The framework will also include a functionality based on machine-learning techniques to detect incorrect/malicious entry of data by certain users. Also, to further reduce the one-time manual entry of labels:
(1) we utilize the well-known Wi-Fi-location databases such as WiGLE~\cite{wiglewifidatabase} to obtain GPS coordinates and then location labels from Google Maps; (2) the newer Android OS, Android 6.0, provides venue name (such as `San Francisco Airport') if published by access point using \textit{ScanResult.venueName} attribute.
\textit{Other devices acquire these labels automatically through collaboration when they visit those locations.}

The features other than the ``list of Wi-Fi Access Points (APs)'' (which we call `additional features') are used to further distinguish different regions within the range covered by a single/set of APs. For example, within the same AP range, it is possible to have two rooms with different sound levels such as a meeting room and a lobby. Because of the advancements in smart phone sensor technologies, some phones now have pressure and temperature sensors, which can be valuable additions to our feature list. We call a tuple of location features and corresponding location label as an \textit{entry}.
The method of populating this database with new entries is outlined below. Our solution needs Wi-Fi to be ON as it relies on Wi-Fi scan results. The application utilizes system scans when the device is not connected to any Wi-Fi network. If the device is connected to a Wi-Fi network, the OS does not initiate scans, so the app initiates the scans. In this case, scans are requested only during the transitions from active to inactive periods detected using accelerometer data and dead reckoning
to conserve energy. These transitions convey that the user has settled down at some place \jj(for some amount of time)\jj after temporary movements. After receiving the scan results, the app checks to see if it is similar to any of the entries in its database. We define a metric called \textit{similarity measure} between two lists of APs to identify how significantly the two regions are similar to each other. Let us denote the list~A of access points obtained at time $t=a$ as $[AP^{(a)}_1,AP^{(a)}_2,\dotsc,AP^{(a)}_m]$ with each term indicating the signal strength (in $W$) of that particular AP (identified using MAC address). Let list~B at time $t=b$ be $[AP^{(b)}_1,AP^{(b)}_2,\dotsc,AP^{(b)}_n]$. We make use of \textit{cosine similarity} to define the similarity measure between these two lists as,
\begin{equation}\label{eq:similarity}
sim= \frac{\sum_{i=1}^{m}\sum_{j=1}^{n}AP_i^{(a)}AP_j^{(b)}\delta_{ij}}{\sqrt[]{\sum_{i=1}^{m}{AP_i^{(a)}}^2} \sqrt[]{\sum_{j=1}^{n}{AP_j^{(b)}}^2}},
\end{equation}where $\delta_{ij} = 1$ if $AP_i=AP_j$ (i.e., their MAC addresses are same), else $0$. If the similarity measure is found to be lower than a certain threshold value ($sim < sim_{th}$) (new location) or if $sim \geq sim_{th}$, but at least one of the location features is different compared to the entries in database, the app initiates a location request (explained in later sections) with the location features collected at that moment. These location features include additional sensor features which are averaged over a small time window (to account for temporal fluctuations), in addition to Wi-Fi AP list. The requester upon receiving the location label distribution (as a response to the request) considers the location label with the highest probability but also with a value beyond a certain threshold (e.g., 0.5) as the correct location label. A new entry is then made to the database using those features and that location label. 
We would like to reiterate that sampling of additional sensor data happens only in the above mentioned scenario (i.e., not continuously), thereby reducing power consumption. Moreover, this information is stored in the device only (not transferred to cloud) thereby posing no privacy problems too.


If no response is returned to the request (i.e., none of the devices know about that location) or if the obtained location label does not meet the above requirements (e.g., too many malicious or privacy concerned users), \jj two situations are possible: (1)\jj the application waits to see if the location is a \textit{significant} location for the user. If the list of Wi-Fi APs remains the same (i.e., $sim > sim_{th}$) for a duration greater than a threshold value ($t_{th}$ = 2 hours, for example), the application recognizes the location as a significant location for the user and prompts at $t = t_{th}$ to enter manually the publicly-known label of that location (this happens only once per location) which is processed by the ontology framework. 
There is also another advantage to waiting for $t = t_{th}$ before asking to manually enter the label--- the same location can be significant to many users, if at least one of them has entered it manually, rest of them can simply obtain it in the process of collaboration. 
Since most of the places are significant to some or other user (who are in fact natives of those places), we believe this is a good approach to one-time manually label them by the natives. 


Finally, we can see that the database in each device will almost saturate after some time. This is because each user is usually associated with only a few places for most of the time. This drastically reduces the need for collaboration until the user visits new places. Hence, for most of the time, the location information is readily available for use in location-based services. We also note that there is a trade-off between the similarity threshold (as in~\eqref{eq:similarity}) and the size of database. If the threshold is more, granularity increases but there are more entries in the database and vice-versa. 

\textbf{Protecting Requester/Provider Privacy: }
In this section, we describe how the requester and providers privacy is preserved (partly). Rest is mentioned in next section.  

\begin{figure}
\begin{center}
\includegraphics[width=3.4in]{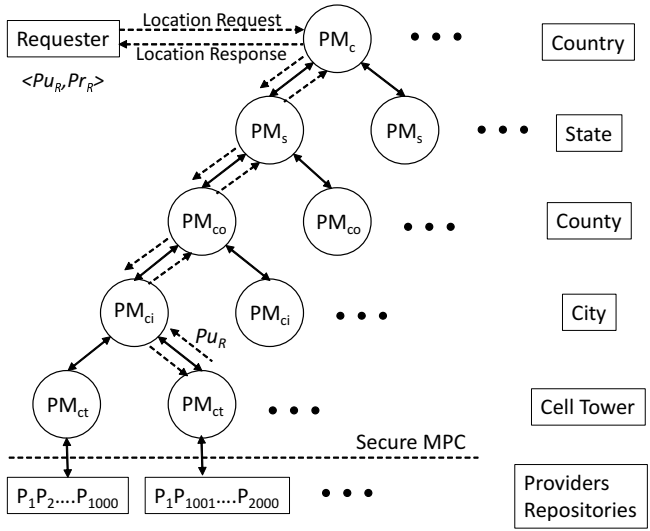}
\end{center}
\vspace{-0.1in}
\caption{An overlay ToR network consisting of Phone Masters~(PMs) (which are also smartphones) at different levels and other user devices. The path taken by an example location request is shown in dashed arrows.}\label{fig:pm_network}
\vspace{-0.2in}
\end{figure}

\textit{1) Area Level Privacy:} 
We will first introduce the concept of ``Phone Masters'' (\textit{PMs}). As shown in Fig.~\ref{fig:pm_network}, PMs are present at different levels forming an overlay network on top of TCP/IP: \textit{Country, State, County, City, Cell Tower}. This can be extended further down (e.g., Access Point level) but we have stopped at Cell Tower level in this paper because cell tower IDs are readily available and a mobile phone can easily find the cell tower ID of its registered cell tower. There are multiple PMs for each level to avoid problems akin to centralized solutions and also for load balancing. To enable uniform load balancing for all PMs, the number of duplications is higher at top levels (e.g., Country level) than at bottom levels (e.g., Cell Tower level). Some of the user devices themselves act as these PMs as we will describe later.
Each device acting as a requester/provider/PM in our solution has an ID that uniquely identifies it. Each PM stores the IDs of its children in its repository similar to a Domain Name Server~(DNS) system. The last level of PMs, namely the Cell Tower PMs~(CTPMs) store the IDs of the location providers that have opted to remain anonymous in that area (we call this database, the repository of that CTPM). A repository is used by CTPM to pick providers to respond to location requests it receives. A provider is given a choice for this area level privacy in the application's privacy preferences. The objective of this privacy option is to anonymize the provider's history of locations to an area level of his/her choice. A provider who wants to remain anonymous within a cell tower region would opt for cell-tower level privacy. His/her ID will be included in the repository of that CTPM. This also means he/she is willing to receive location requests from any requester currently located within that cell-tower area. Similarly, a provider who wants to remain anonymous within a city region would opt for city-level privacy. Since a city has multiple cell towers in general, his/her ID will be included in the repositories of all CTPMs of that city like provider $P_1$ in Fig.~\ref{fig:pm_network}. So s/he will be receiving requests from devices located in any of those cell tower regions. There is an obvious trade off here---higher the privacy level, more the requests and vice-versa.
A requester device with a location request corresponding to a cell tower, first contacts one of the top-level (Country) PMs; the request is then forwarded down to PM of that cell tower using the overlay network of PMs. The CTPM will then pick the appropriate providers from its repository to help answer the request through weighted collaboration. The answer is finally sent back to the requester using the same path. We describe below how this process is anonymous as the above overlay network acts as a ToR network.



\textit{2) ToR Network: }In our scenario, the overlay network of PMs form the \textit{ToR (Onion Routing) network}~\cite{tor} which provides anonymous connection between two parties over a network of nodes (network of PMs in our case).
Since our solution works at application level, we need either a mapping from device IDs to IP addresses or the IP addresses themselves can be used as device IDs. One may be of notion that IP addresses reveal location information. However, IPs do not provide accurate location information, especially for mobile devices. Additionally, existing techniques such as IP spoofing can be used to prevent location information disclosure. A requester conceals its location features (corresponding to a cell tower) in the onion packet with final destination being the PM of that cell tower. Requester also generates a public-private key pair ($Pu_R,Pr_R$) for \textit{each} request and includes the public key, $Pu_R$, along with the onion packet in the request (Fig.~\ref{fig:pm_network}). It sends this request to a topmost level PM \jj (akin to root servers in DNS)\jj and each PM then forwards the request to appropriate children until it reaches the CTPM it is destined to. Once the CTPM receives the request it queries the providers in its repository to find the final location distribution. It then encrypts this distribution with $Pu_R$ and 
sends back to the device from which it received the request originally (each PM does the same), finally reaching the requester. The encryption by CTPM is to ensure that Country PM which knows requester's ID does not know the location information of the requester\jj (whole process similar to recursive lookup in DNS combined with ToR routing)\jj. Moreover, when the providers provide location labels to CTPMs, their location \textit{history} will be known to CTPM (which is less severe than the \textit{current} location in case of requester). This issue can be addressed using secure Multi-Party Computation (MPC) protocols~\cite{Dugan2016ATests} such as Garbled Circuits
or Shamir's secret sharing
between providers and CTPM. An example path taken by a location request is shown in Fig.~\ref{fig:pm_network}.

\textbf{Final Distribution Generation: }
A CTPM upon receiving the location request first picks $j$ devices (providers) at random from its repository to answer the location request. Algorithm~\ref{algo:dist_gen} summarizes the procedure followed by each of the $j$ devices to generate their distribution tailored to their privacy levels. It consists of (i) generating correct (without adding noise) distribution by applying two-step classifier on their databases, 
(ii) introducing noise to the distribution by adding a certain number of random labels and then a random noise to the probabilities. (iii) picking the top $k$ labels. The k-label responses from all such providers are then fused by the CTPM using weighted average fusion to generate final distribution.

\begin{algorithm}[t!]
\SetAlgoNoLine
{\fontsize{8}{8}\selectfont
\caption{Location Distribution Generation~\textit{(LDG)}}\label{algo:dist_gen}
\KwIn{Database $D: W,F_1,F_2,...F_n, L$ (size m); Input $I: w,f_1,f_2,...f_n$; Privacy levels: $p_1, p_2$; $k, (r_1,r_2)$}
\KwOut{Location Label Distribution: $V$ (length $k$) }
\nl $M = S_1 = R = V = \emptyset$; $sum = 0$ \\
\nl \lForEach {$w_i \in W$}{
\uIf {$sim(w_i,w) != 0$}{
$M \leftarrow M \cup index(w_i)$ \\
$S_1 \leftarrow S_1 \cup \{S_1[l_i] = sim(w_i,w)\}$}}
\vspace{-0.1in}
\nl $S_2 \leftarrow NFM(D[M],I)$; $S_1 \leftarrow Normalize(S_1)$ \\
\nl $S \leftarrow WeightedAvgFusion(\{S_1,S_2\},\{r_1,r_2\})$ \\
\nl R $\leftarrow$ Generate $p_1$ random labels \& Initialize probs to 0; $V \leftarrow S \cup R$\\
\nl \lForEach {$l_i \in V$}{$V[l_i] \leftarrow V[l_i] + N(0,p_2)$} 
\nl Scale and Normalize $V$ \\
\nl Sort $V$ by probabilities in decreasing order \\
\vspace{-0.05in}
\nl \Return $V = \{V_1,V_2,...V_k\}$
}
\end{algorithm}
\setlength{\textfloatsep}{3pt}

\textit{1) Two-step Classification:}\label{sec:two-step}
Each of the $j$ providers picked by CTPM runs this algorithm to return correct distribution over their database labels given the location features. It consists of the following steps: \textit{(1)} Each device runs a similarity check (using~\eqref{eq:similarity}) between the input Wi-Fi AP list and the lists(entries) in its database. 
\textit{(2)} If the similarity returned is 0 for all entries, it returns  ``NA'' indicating ``I don't know''. We note that there may be a considerable number of such devices (out of $j$) because of two reasons: first, ``Area-level privacy'' preferences mentioned above (e.g., the device might not have visited any place in that cell tower area but its ID is listed in that CTPM's repository due to its city-level privacy requirements) and second, the fact that many of them have not visited \textit{that particular} location in the past even though they have visited other locations in that cell tower area. 
\textit{(3a)} If not, an NFM classifier (described below, also as Algorithm~\ref{algo:nfm_algo}) is trained over the additional features (other than Wi-Fi) belonging to these entries that have non-zero similarity measure. Then the additional features in the input are queried against this classifier to generate a distribution over the labels corresponding to these entries. \textit{(3b)} A second probability distribution is generated over the same set of labels by normalizing the similarity measures. \textit{(3c)} The distributions in \textit{(3a), (3b)} (thus two-step) are fused using weighted fusion. Weights, ($r_1,r_2$), can be determined empirically or through prior knowledge.
\textit{(3d)} Remaining labels in the database are also added to the distribution in \textit{(3c)} with probability zero which is the output of two-step classification.
The idea behind considering devices with non-zero similarity measure \jj (which means at least one AP is common between the requester and provider devices)\jj is to identify devices which have at least some knowledge about the query location (e.g., nearby locations with a common Wi-Fi AP).

\begin{algorithm}[t!]
\SetAlgoNoLine
{\fontsize{8}{8}\selectfont
\caption{NFM Classifier~\textit{(NFM)}}\label{algo:nfm_algo}
\KwIn{Training Set $T: F_1,F_2,...F_n, L$ (size m); Input $I: f_1,f_2,...f_n$}
\KwOut{Location Label Distribution: $V$ (length m)}
\nl $count[m] = sum = 0;V\{m\}=\emptyset$ \\
\nl \lForEach {$feature~vector~F_i$}{
 \uIf{$class(F_i) == numeric$}{
 \text{Form Categories $F_{icat}$} $\ni$ boundaries = means of adjacent entries}}
 \vspace{-0.1in}
\nl \lFor{$i~in~1:m$}{
 \lFor {$j~in~1:n$}{
 \uIf {$f_j \in F_{jcat}$} 
 {$count[i]=count[i]+1$;\,$sum = sum+count[i]$}}}
  \vspace{-0.2in}
\nl \lForEach {$l_i \in L_n$}{
$V[l_i] = count[i]/sum$}
\nl \Return $V$
}
\end{algorithm}
\setlength{\textfloatsep}{3pt}

\textit{2) Number of Feature Matches (NFM) Classifier:}\label{sec:nfm} 
We developed a classifier based on the number of feature matches (Algorithm~\ref{algo:nfm_algo}) as we found the existing techniques not well-suited/give low accuracy. Even though Multinominal Logistic Regression (MLR) suits our problem, NFM is found to perform better than MLR. 
NFM \jj is developed keeping in mind that the training data is small - \jj takes only those entries with non-zero similarity with the input Wi-Fi list as training data. These entries correspond to adjacent areas such as rooms in a building. Given the training data (consisting of `additional' features and labels) and query features, the classifier generates a distribution over the training labels using the following steps: (1) Maintain a count for each entry/label (2) For each `additional' feature column generate categories corresponding to feature values such that the category boundaries are at the mean of two adjacent feature values (e.g., for $(1,5,6)$ feature values, three categories are $(-\infty,3),(3,5.5),(5.5,\infty)$) (3) For each entry update the count as follows - for each feature value, if the query feature value falls in its category, increment the count. (4) Normalize the counts for all entries to generate a probability distribution over the labels.  

\textit{3) Distribution Perturbation:}\label{sec:dist_pert} 
The distribution generated by the two-step classification procedure is enlarged, perturbed and then truncated before sending back to CTPM. First, $p_1$ additional random labels are added to the list with probability values set to zero. These labels are sourced randomly from a database containing location labels belonging to the provider's \textit{Area-level Privacy} region. This database which can be built one-time from Google Maps is stored locally in each device. Second, random noise, $n\sim \mathcal{N}(0,p_2)$ is added to all the labels in the distribution which is then scaled and normalized. Both $p_1$ and $p_2$ are set by the privacy level needed by the devices. The higher these variables are, the more noise is added to the distribution. It is clear to see that for high values of $p_2$, we obtain a uniform distribution. The app then selects the top $k$ labels with highest probabilities and sends them back to the CTPM. Until the desired number ($l$) of non-NA responses are obtained, the CTPM repeats the process, each time, by doubling the number of devices ($j_{new} = 2j_{old}$) picked from the remaining providers. 
\begin{figure*}[t!]
        \centering   
           \begin{subfigure}[b]{0.33\textwidth}  
            \centering 
            \includegraphics[width=1\textwidth,height=2.5in]{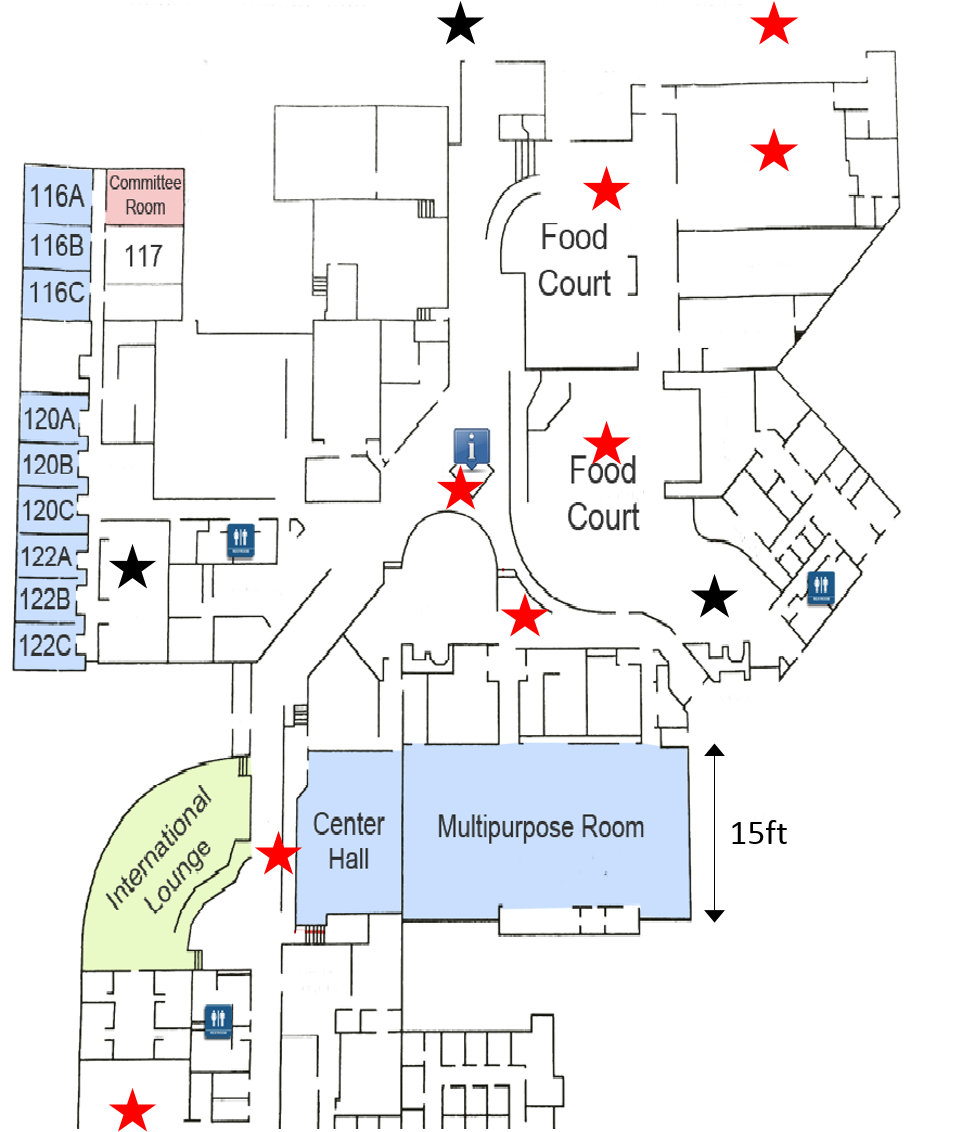}
            \caption{}
            \label{fig:busch-student-center-final}
        \end{subfigure}%
~
        \begin{subfigure}[b]{0.33\textwidth}   
            \centering 
            \includegraphics[width=1\textwidth]{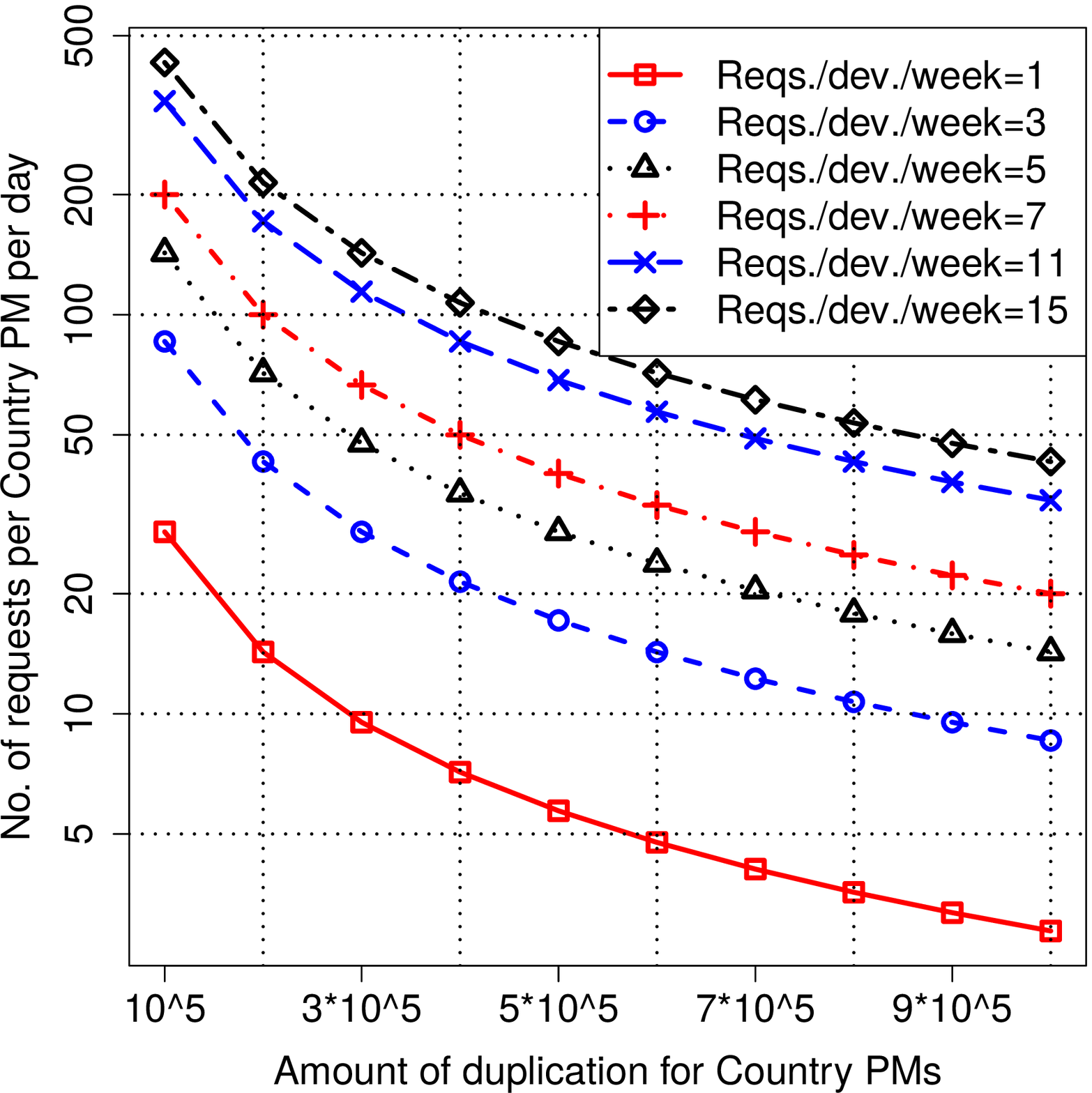}
            \caption{}
            \label{fig:numreqs-vs-dup-var-numreqs}
        \end{subfigure}%
~        
           \begin{subfigure}[b]{0.33\textwidth}
        		\centering
        		\includegraphics[width=1\textwidth]{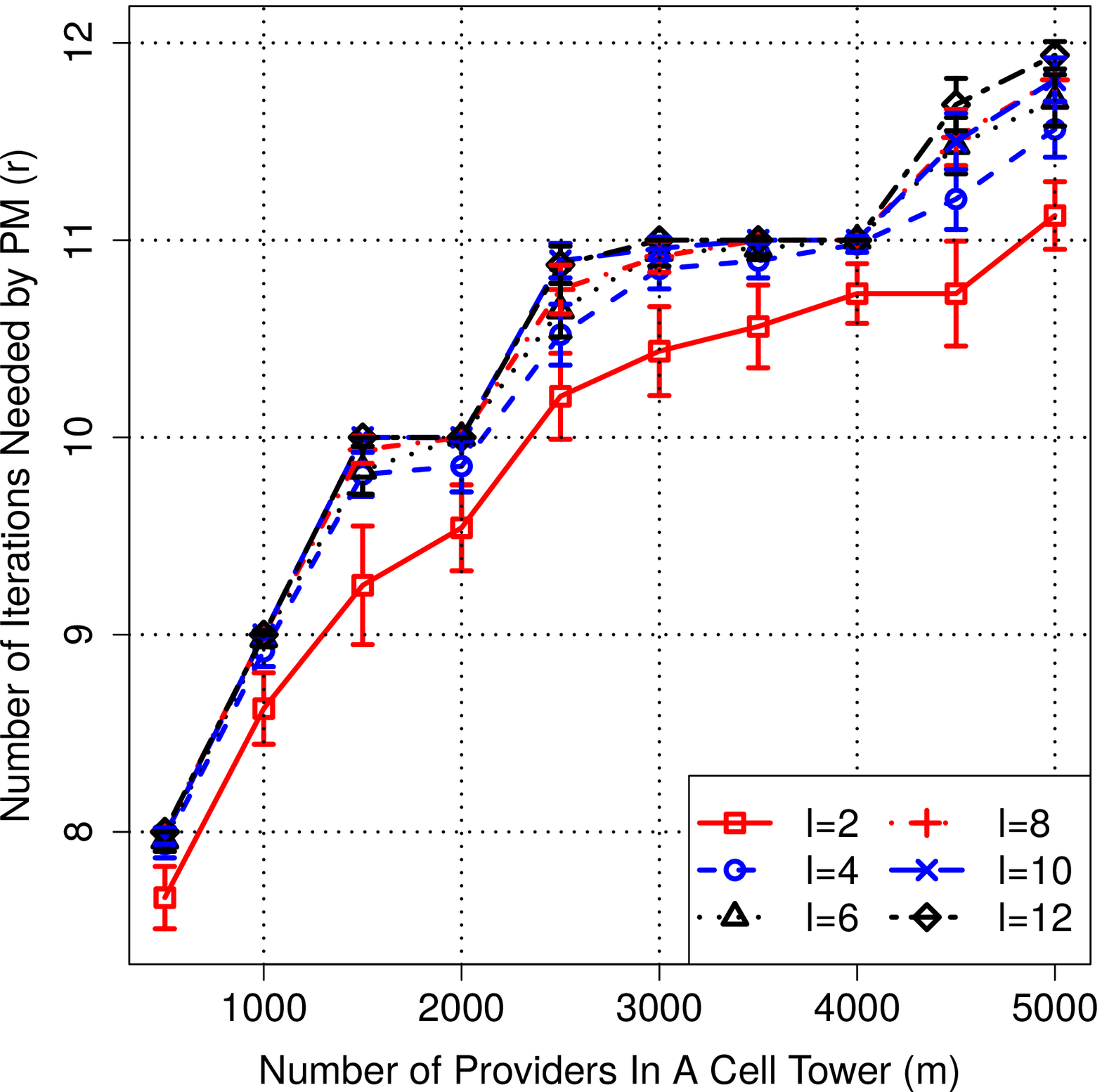}
        		\caption{}
        		\label{fig:r-vs-m-var-l}
        	\end{subfigure}
        \vspace{-0.2in}
        \caption{\label{fig:app} (a) Floor map of Student Center showing data collection spots. (b) Number of requests received by Country PM per day vs. PM duplication and new requests made by requester per week (c) Number of iterations by CTPM vs. size of its repository and number of collaborators needed. }
        \vspace{-0.15in}
\end{figure*}

\textit{4) Weighted Information Fusion:}
The $l$ non-NA responses are fused using weighted average with the weights as the utility values of the devices to generate the final distribution, which is sent back to the requester using the same path as the request. Here the utility, $U$, consists of two components: a noise component ($U_n$) and a time component ($U_t$): $U = U_n \times U_t$. $U_n$ 
estimates a device's utility in terms of providing useful information (considering noise/privacy/maliciousness). Devices acquire these values based on past feedback they received from CTPMs, similar to aggregating global reputation scores from local feedback in a peer-to-peer system~\cite{powertrust2007}.
%
$U_t$, on the other hand, indicates that more weight has to be given to the latest entries as the Wi-Fi APs might have changed over time. 
%
The probabilities in the final distribution give confidence estimate of the respective labels. This is especially helpful in partial coverage areas (e.g., only 30\% of the rooms are labeled). 
\section{Performance Evaluation}\label{sec:perf_eval}

We first explain our experimental setup,
then present theoretical results followed by experimental results on real location data in terms of privacy, accuracy, collaboration metrics. 

  

\textbf{Experimental Setup: }Since our solution is pervasive (i.e., works across buildings), we collected data at two separate regions (totaling fifteen locations/buildings) separated by a cell tower distance.
Four of these locations \jj (top place-marks in the map)\jj are located in a campus region while \jj the bottom place-marks\jj the rest are located in a downtown region. We used seven Android mobile devices in our experiments out of which four are phones and three are tablets. Within each of these locations, a person with the app installed on his device moved around these regions. Whenever the app detects that the received Wi-Fi scan results are dissimilar (we used a threshold value of $0.05$, see~\eqref{eq:similarity}) with the entries in the phone's database (initially database is empty), the application recorded the following location features---(1) list of Wi-Fi Access Points (APs) along with SSID, BSSID (MAC Address) and signal strengths (we limited it to 15), (2) sound level, (3) cell tower ID, (4) Location Area Code (LAC), (5) cell signal strength (in $\mathrm{dBm}$). One location out of the 15, is a student center shown in Fig.~\ref{fig:busch-student-center-final}. Both red and black stars indicate the places where the app recorded location features (any two of these places have a Wi-Fi similarity of less than $0.05$).

Each device is able to record on an average of 50 entries corresponding to different rooms/places \textit{in} and \textit{around} the above fifteen buildings (this is not exhaustive coverage, as it represents a typical user's usually visited places). In case of large buildings (like campus buildings, library), multiple entries are recorded (as the user spends more time and explores more rooms/places in those buildings)\jj corresponding to different rooms in the same building\jj, while in case of small buildings with one or two small rooms (like Starbucks), only one entry is recorded. As we can see, each phone is not limited to a single building as in indoor localization solutions but collects data from different places it visits.
We have made the following observations during our experiments: Granularity (1) can be increased or decreased depending on the similarity threshold. More is this value, we have more granularity and vice versa; however other location features could limit this granularity if they do not exhibit significant changes in their values (due to two-step classification). (2) depends on the device's ability to capture the surrounding APs. For the same similarity measure, a device with advanced Wi-Fi radio chip is able to distinguish well between the regions than a less capable device. We referred to this as one of the sources of sensor noise earlier. 
(3) depends on the density of APs in the region. In an urban setting where there is more density of Wi-Fi APs, more granularity is possible compared to a sub-urban or rural area with less density.

Sound level is recorded as the \textit{median} value of the maximum amplitude (\textit{getMaxAmplitude()} in Android), which indicates the loudness of the sound, of three consecutive one-minute intervals. Cell signal strength is recorded as the \textit{mean} signal strength over two-minute interval. \textit{We believe these intervals are sufficient enough to account for fluctuations.} These two features are considered numeric values while the cell tower ID and LAC as categorical values. For each of these features in the database, we entered the label manually. We would like to clarify that this is for evaluation purposes only. In actual deployment, the app would first attempt to acquire this label through collaboration via the ToR network of PMs. 

\begin{figure*}[t!]
        \centering   
           \begin{subfigure}[b]{0.33\textwidth}
        		\centering
        		\includegraphics[width=1\textwidth]{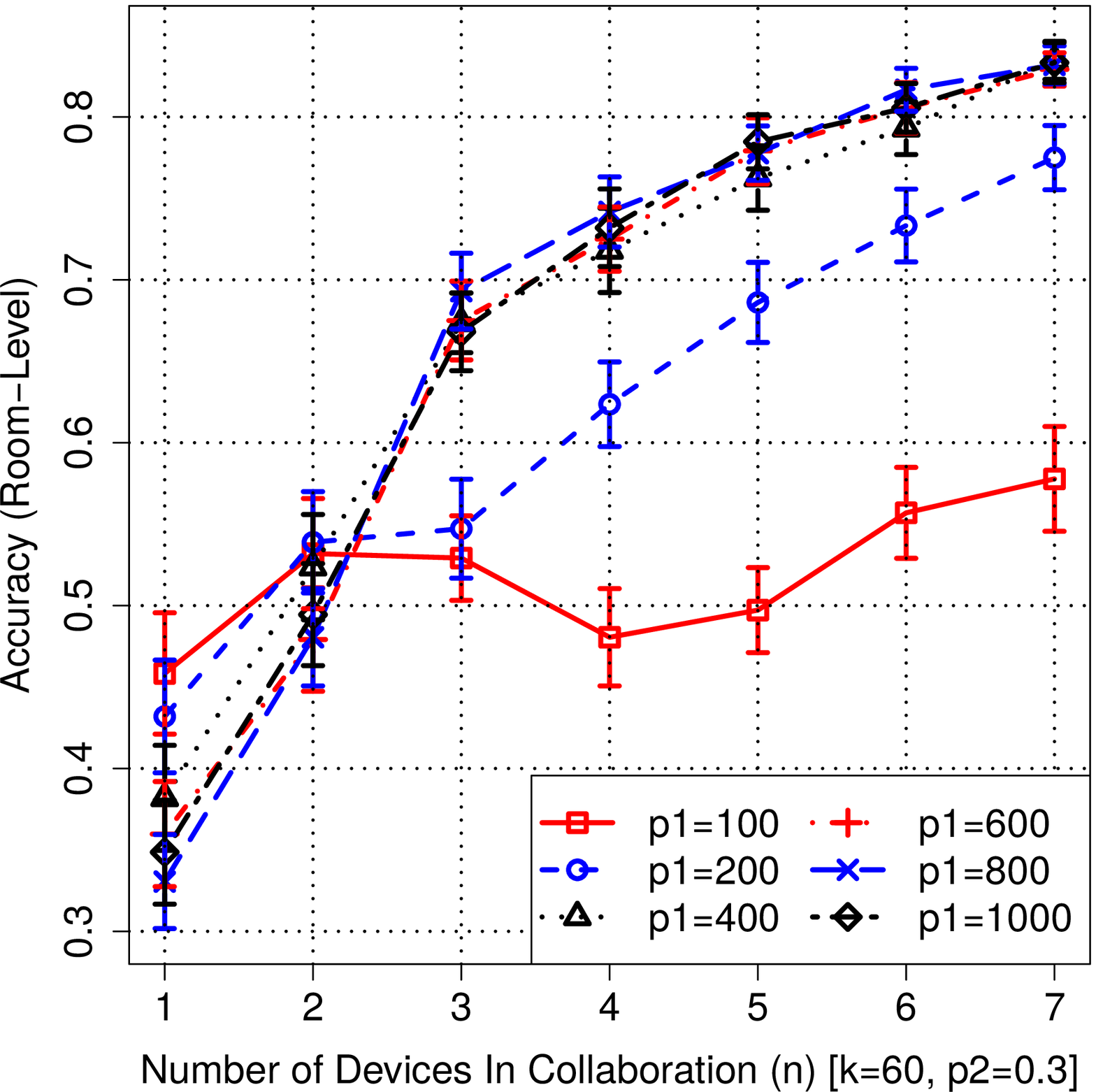}
        		\caption{}
        		\label{fig:a-vs-n-var-p1}
        	\end{subfigure}%
~
        \begin{subfigure}[b]{0.33\textwidth}  
            \centering 
            \includegraphics[width=1\textwidth]{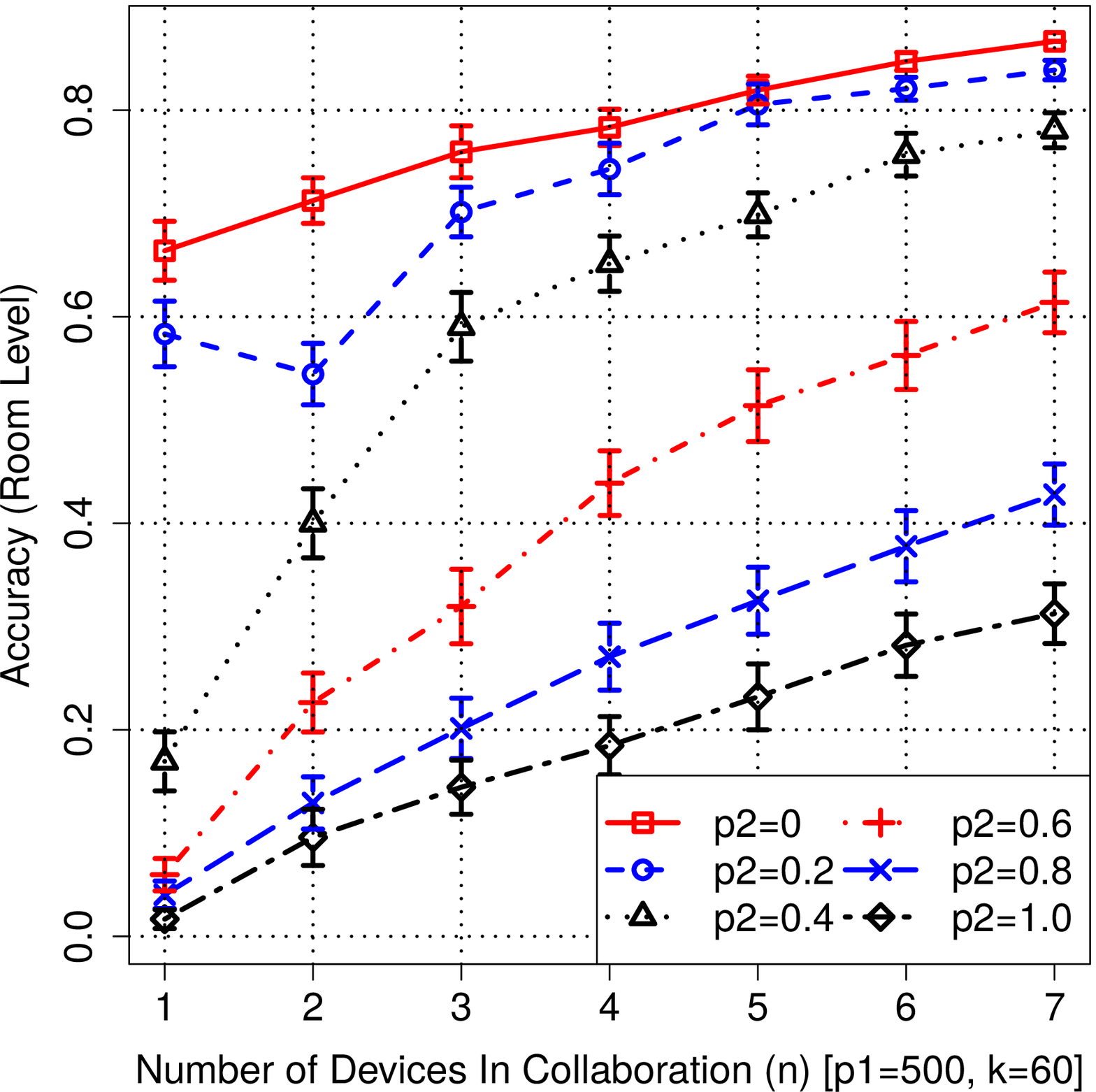}
            \caption{}
            \label{fig:a-vs-n-var-p2}
        \end{subfigure}%
~
        \begin{subfigure}[b]{0.33\textwidth}   
            \centering 
            \includegraphics[width=1\textwidth]{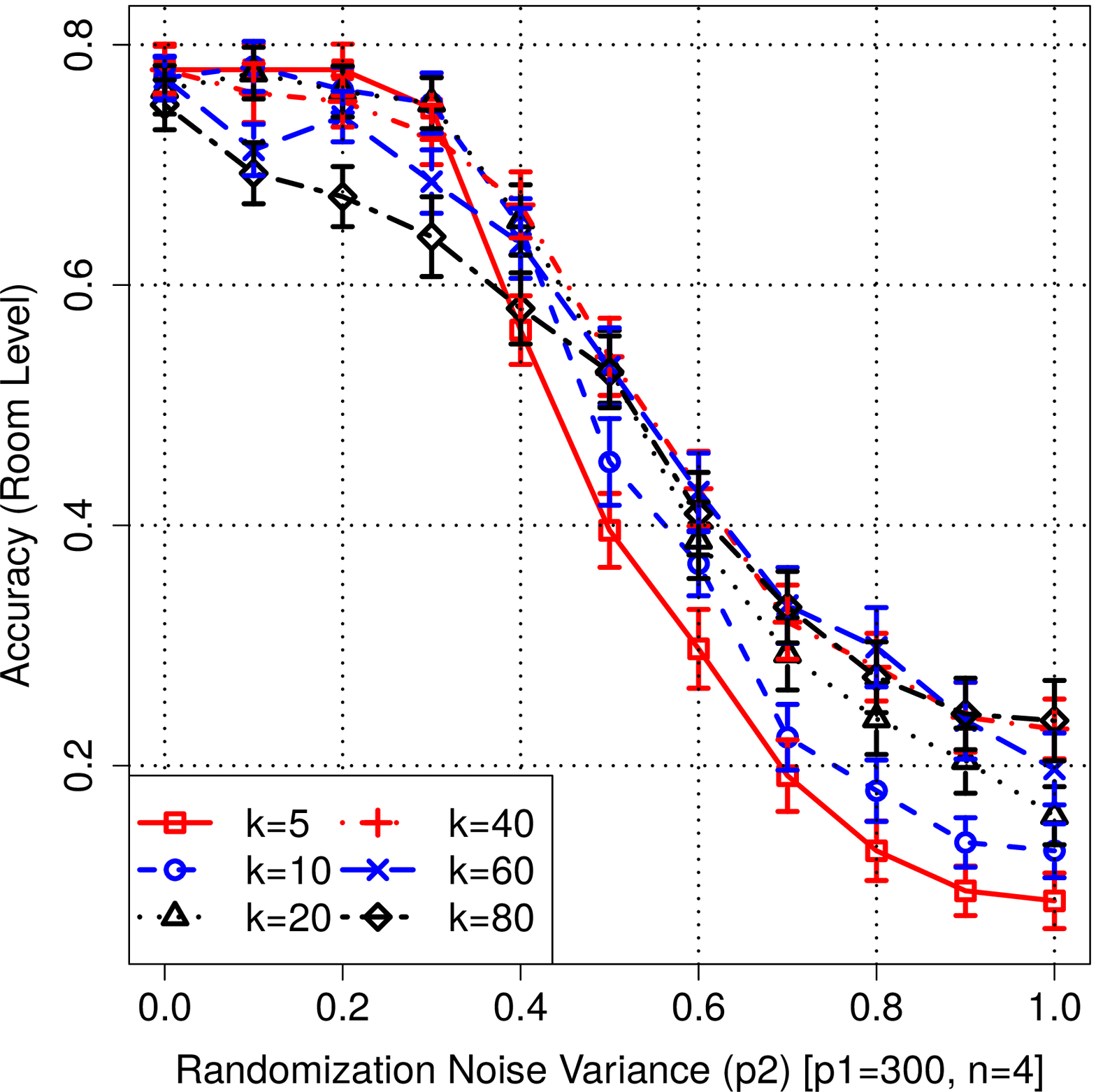}
            \caption{}
            \label{fig:a-vs-p2-var-k}
        \end{subfigure}
        
        \begin{subfigure}[b]{0.33\textwidth}
        		\centering
        		\includegraphics[width=1\textwidth]{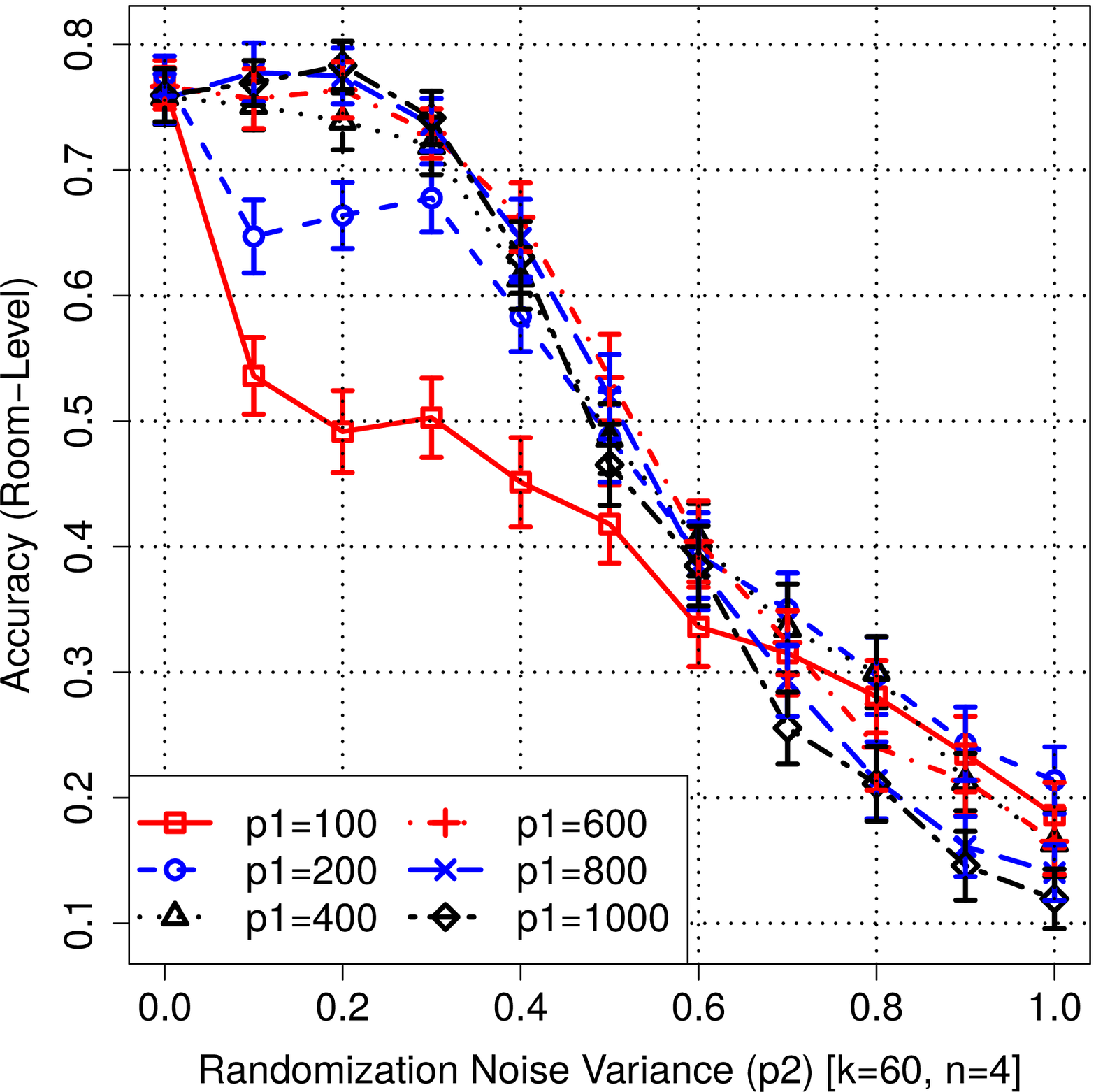}
        		\caption{}
        		\label{fig:a-vs-p2-var-p1}
        	\end{subfigure}%
~
        \begin{subfigure}[b]{0.33\textwidth}  
            \centering 
            \includegraphics[width=1\textwidth]{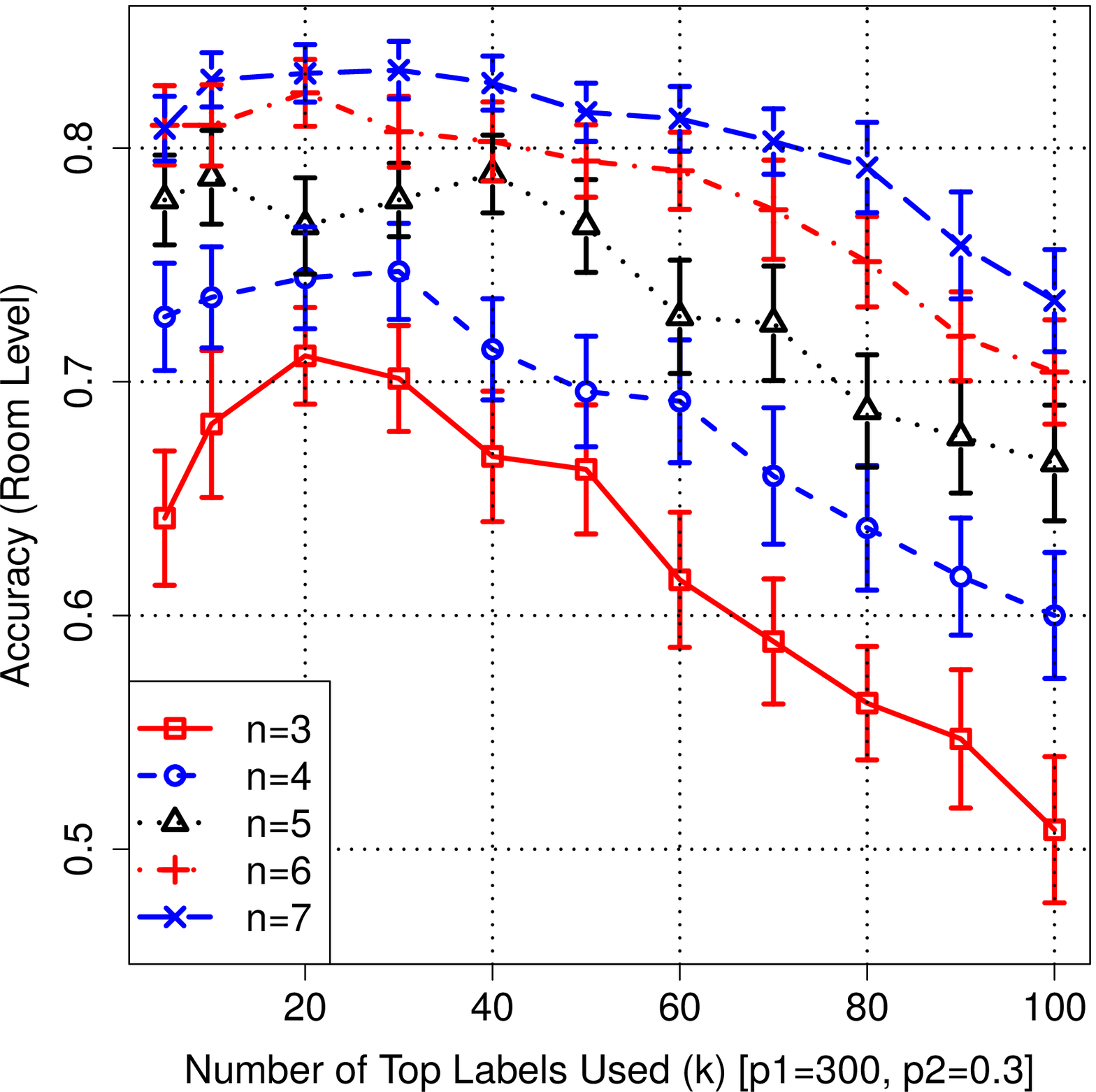}
            \caption{}
            \label{fig:a-vs-k-var-n}
        \end{subfigure}%
~
        \begin{subfigure}[b]{0.33\textwidth}   
            \centering 
            \includegraphics[width=1\textwidth]{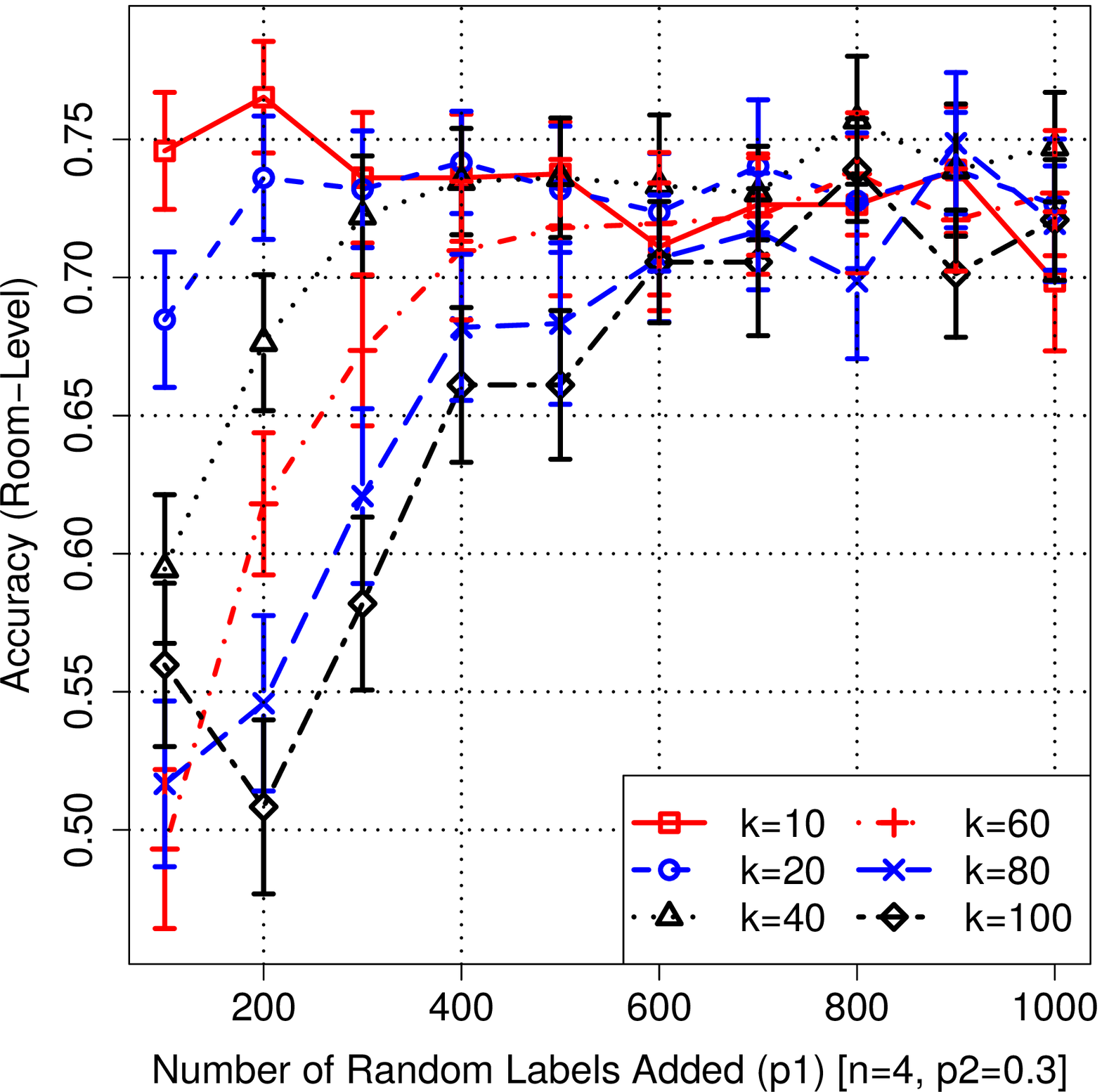}
            \caption{}
            \label{fig:a-vs-p1-var-k}
        \end{subfigure}
        \vspace{-0.2in}
        \caption{\label{fig:exp_results} Experimental results showing how accuracy varies for different values of design parameters---$n, p_2, p_1, k$.}
        \vspace{-0.15in}
\end{figure*}

%
\textbf{Theoretical Results.} Currently there are about 200 million smartphone users~\cite{HowEMarketer} out of which we assumed $10\%=20e6=n_R$ to use our solution considering progressive adoption. Figure~\ref{fig:numreqs-vs-dup-var-numreqs} shows the number of location requests received by a Country PM per day ($nreq_{PM_c}$) vs. number of Country PMs ($n_{PM_c}$) and the number of requests generated per week per requester ($nreq_{R}$) assuming the following relation: $nreq_{PM_c}=n_R.nreq_{R}/(7.n_{PM_c})$. We note that only in the beginning of app install there will be high number of requests (such as 15 per week), but from second week onwards it will be $< 5$ as many places are already known. \textit{This is because our app tries to localize only when the user's location is stable for a certain amount of time.} We can observe that even for $n_{PM_c} = 1e5$, the maximum $nreq_{PM_c} \approx 400$ which is still manageable. For most of the later weeks, $nreq_{PM_c} < 100$, considering $3$ \textit{new} location requests per requester per week. These results are also applicable to PMs at other levels due to uniform number of PMs at all levels.
Figure~\ref{fig:r-vs-m-var-l} shows the number of iterations, $r$, to be done by a CTPM to obtain the desired number of collaborators, $l$, as the size of its provider repository, $m$, varies from 500 to 5000. We remind that the collaborators are the devices that present non-NA response to CTPM.
Besides showing $r$ needed for a given $l,m$, we can also observe that $r$ does not change much as the number of collaborators increases, which shows that it is best to choose a high number of collaborators since the number of extra iterations required is low. All the results plotted in this paper with confidence intervals are averaged over 48 runs, which guarantees statistical relevance. In each of the 48 runs we randomized the $n$ devices picked (out of 7) for collaboration. For example, if $n=4$, $7~choose~4=35$ combinations are possible and $48(>35)$.

\textbf{Experimental Results: }
We considered two accuracy levels: exact/fine (approximately room level) and coarse (building level). \textit{Here we would like to clarify the difference between accuracy and granularity.} Granularity of our system is set by the similarity threshold used~\eqref{eq:similarity}, which we have set to $0.05$ to get approximately room level granularity (around 5 meters). This number can be varied to get room-level or sub-room level granularity.
Accuracy on the other hand is determined by the percentage of test cases that are classified correctly to be the same region/room label. If the algorithm is able to detect the exact room/place in a building, we call it room-level accurate (which is also building-level accurate), while if it detects it to be another room/region within the same building, we call it building-level accurate (but not room-level accurate). If the classification is to a place in another building, it is neither-level accurate. We remind that building-level accuracy is sufficient for coarse location-based services (reason for considering this) and room-level accuracy can be used, for example, to enforce security policies as mentioned earlier. 
To test the algorithm, we collected location features 
at 15 places (out of 50 places in training data), which is used as test data. Black stars in Fig.~\ref{fig:busch-student-center-final} show test data points for Student Center. For each of these test places, a location label is found from collaboration as per $n$ (number of collaborators = $l$), $k$ (number of top labels used in the fusion process), $p_1$ (number of random labels added to the distributions), $p_2$ (randomization noise) values. 

\begin{figure*}[t!]
        \centering  
        \begin{subfigure}[b]{0.33\textwidth}  
            \centering 
            \includegraphics[width=1\textwidth]{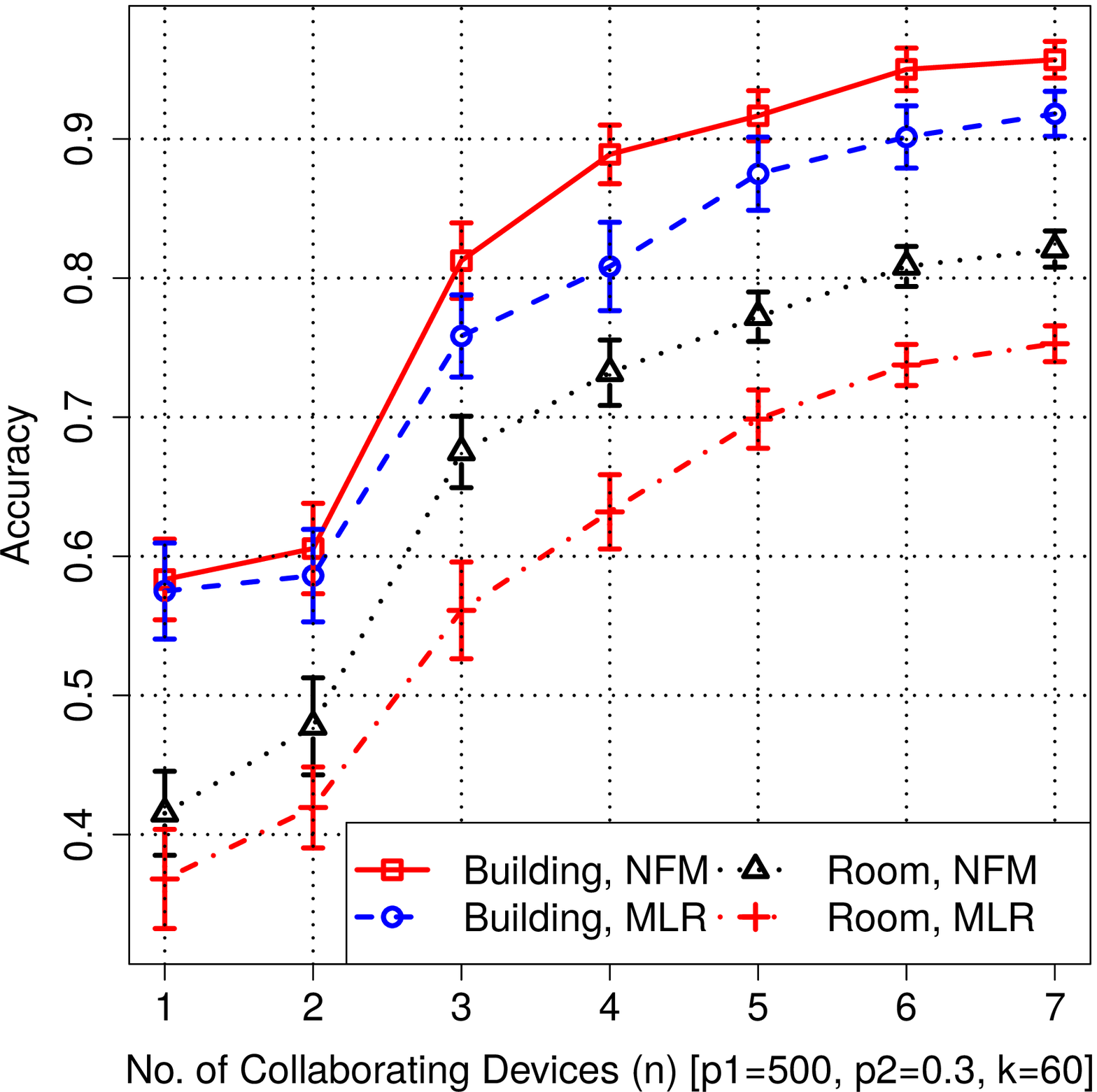}
            \caption{}
            \label{fig:a-vs-n-var-algo}
        \end{subfigure}%
~
        \begin{subfigure}[b]{0.33\textwidth}
        		\centering
        		\includegraphics[width=1\textwidth]{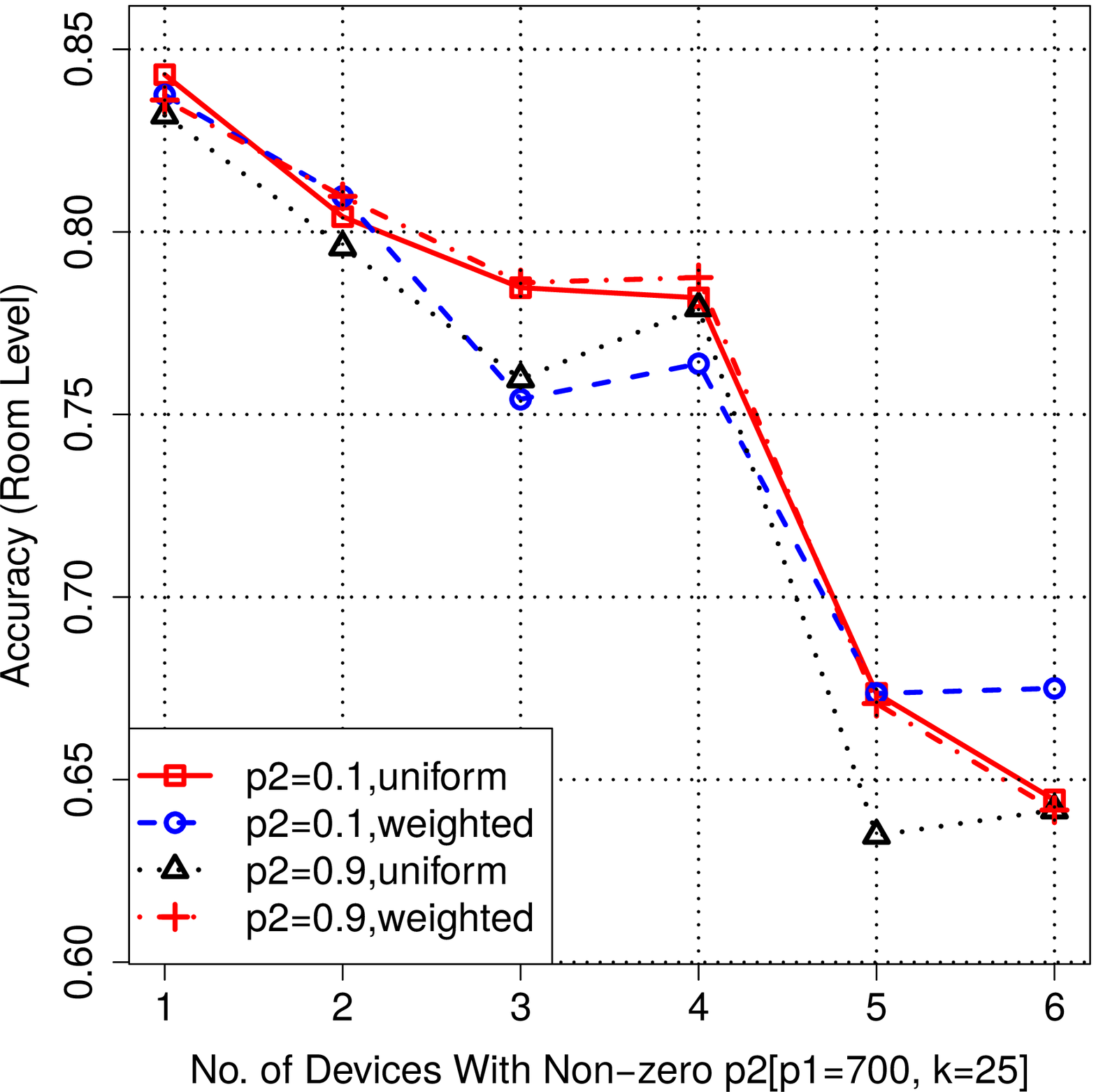}
        		\caption{}
        		\label{fig:a-vs-n-var-p2_utils}
        	\end{subfigure}%
~           
        \begin{subfigure}[b]{0.32\textwidth}  
            \centering 
            \includegraphics[width=1\textwidth]{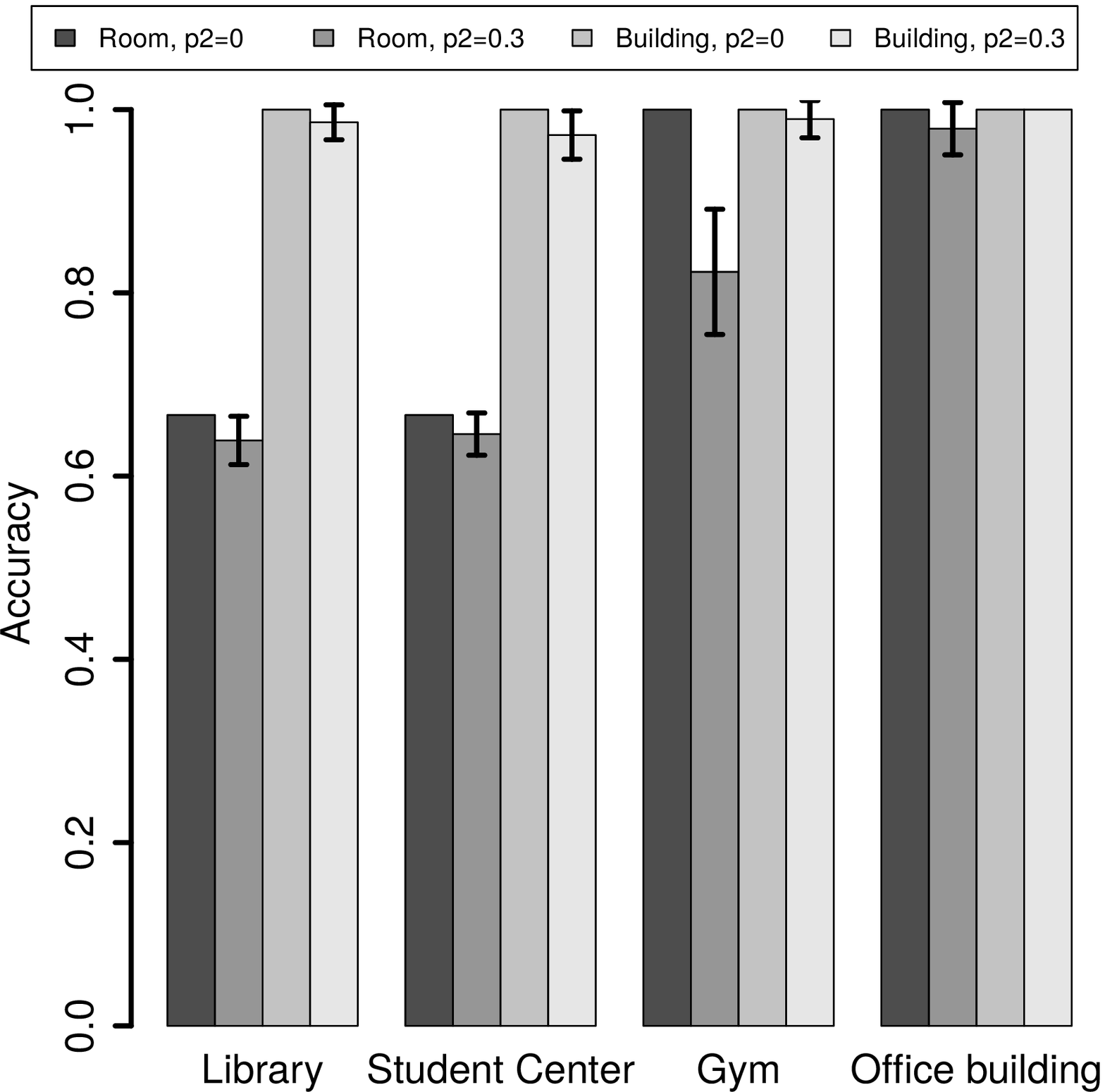}
            \caption{}
            \label{fig:a-vs-building-var-p2}
        \end{subfigure}
        \vspace{-0.2in}
        \caption{\label{fig:result_set_3} (a) Room- and Building-level accuracies with NFM and MLR classifiers vs. number of collaborating devices ($n$) (b) Effect of weighting and $p_2$ on accuracy. (c) Room- and Building-level accuracies for different buildings and $p_2$ values}
        \vspace{-0.2in}
\end{figure*}

\underline{Privacy evaluation.} In the following results, we evaluate the privacy features of our framework (excluding the ToR/PM network)
by studying how accuracy varies with key privacy parameters, $k, p_1,p_2$ and $n$, number of collaborators. We used equal weights for all $n$ devices in collaboration and plot only room-level accuracy as it is more fine-grained. Similarly we used NFM classifier as it is more accurate than MLR classifier (shown later).

Figure~\ref{fig:a-vs-n-var-p1} shows the accuracy vs. the number of devices in collaboration for different cases of $p_1$. We can note the general trend that as $n$ increases accuracy increases. Trend for $p_1$ is, however, interesting. We can see that if $n=1$ lower $p_1$ is better, and for $n>2$ higher $p_1$ is better. This is because, for $n=1$, more $p_1$ leads to more noise in terms of number of random labels occupying the top $k$ slots, and this noise is not removed due to lack of collaboration. However, the same noise is beneficial for $n>2$ as the noise added is disjoint across devices canceling out in the fusion process, thus making the correct labels stand out (i.e., the competition among database labels is reduced). Note that this behavior is slightly against intuition.

Figure~\ref{fig:a-vs-n-var-p2} shows the accuracy as $n$ is increased for different cases of $p_2$. We can appreciate the expected behavior w.r.t. $n$ and $p_2$. Another interesting observation is the reduction in uncertainty (confidence intervals) as the number of devices increases. This result proves that collaboration not only enhances accuracy but also improves the confidence of the result. We can draw interesting inferences from this figure---e.g., to obtain 60\% accuracy we need either three devices with average $p_2 = 0.4$ or seven with average $p_2 = 0.6$.

Figure~\ref{fig:a-vs-p2-var-k} shows the accuracy as $p_2$ is increased for different values of $k$ (number of top labels used in fusion). We can see the expected trend of reduction in accuracy as $p_2$ is increased. When the noise ($p_2$) added is low, using more labels ($k$) increases competition among the database labels resulting in accuracy reduction. This explains the low accuracy for high $k$ when $p_2$ is low. However, if $p_2$ is higher most of the top labels are random labels, so taking more of them during fusion increases the chances of finding correct labels in the fusion output (as the random labels are disjoint across devices and hence cancel out). This explains the interesting criss-cross behavior also observed before in Fig.~\ref{fig:a-vs-n-var-p1}.

Figure~\ref{fig:a-vs-p2-var-p1} shows the accuracy versus $p_2$ for different cases of $p_1$. Apart from the general behavior of reduction in accuracy with the increase of $p_2$, we again observe criss-cross behavior w.r.t. $p_1$. When $p_2$ is small, less noise is added to distribution. In this situation, adding fewer random labels means that few of them are present in the top $k$ labels, which in turn leads to more competition among database labels resulting in low accuracy. However, when more noise is added, adding more random labels decreases accuracy as they kick the correct labels out from the top $k$ slots (so small $p_1$ is better). 
Figure~\ref{fig:a-vs-k-var-n} shows the accuracy vs. number of labels used in fusion and $n$. 
Again, we can see that the uncertainty reduces as $n$ increases and vice-versa. We observe a slight increase-decrease behavior with $k$. This is because for small $k$ there is less chance for correct labels to appear in top $k$ slots (reducing accuracy) and for large $k$ there is more competition from database labels again reducing accuracy. 
This means that there is an optimum $k$---this can be seen in the figure and seems to be approximately same for most of the devices equal to 20-30. Again, this behavior is beyond intuition. 

Figure~\ref{fig:a-vs-p1-var-k} shows the accuracy as the number of random labels ($p_1$) added is increased for different values of $k$. 
We can observe that as $p_1$ increases, accuracy increases which is against intuition. This is because, more $p_1$ leads to more noise in terms of number of random labels occupying the top $k$ slots. This noise however is beneficial (for $n>2$) as the noise added is disjoint across devices thereby canceling out in the fusion process, and making the correct labels stand out (i.e., the competition among database labels is reduced).
This particular trend however diminishes as $k$ is decreased (notice $k=10$ curve). This is because the disjoint randomization created by $p_1$ is limited by $k$. Also, for small $k$, there is more chance that the correct labels are not captured in the top $k$ slots reducing accuracy. It is interesting to note that some of these results are against/beyond intuition\jj, such as for higher values of $k$, the accuracy reduces with increasing $k$, etc\jj. From these figures it is possible to obtain optimum values for $k,p_1$. For $k$ as mentioned before (Fig.~\ref{fig:a-vs-k-var-n}) 20-30 is a good number. For $p_1$, consider Fig.~\ref{fig:a-vs-p1-var-k}, assuming a $k$ value of 20 or 40, we can see that $p1>400$ provides good accuracy. Since we need $p_1$ to be greater to preserve privacy, it is best to choose large value for $p_1$ without affecting accuracy. 



\underline{NFM vs. MLR classifiers.} Figure~\ref{fig:a-vs-n-var-algo} 
compares NFM and MLR classifiers as number of devices in collaboration ($n$) increases. We can see that NFM performs at least as good as MLR. Similar result of NFM performing better than (or at least as good as) MLR is observed when the randomization noise variance ($p_2$) is changed (not shown due to space limitation). 

\underline{Room and building level accuracy.} We can also observe in Fig.~\ref{fig:a-vs-n-var-algo}, building-level accuracy is always higher than the room-level accuracy. Both of them increase as $n$ increases and decrease as $p_2$ is increased. The reasons for lower room-level accuracy are: (1) changing Wi-Fi signal strengths, (2) sensor noise---unable to detect some APs; Android OS returning previous scan results as current ones (we have observed this in our experiments), and (3) possibly low amount of training/test data (i.e., more data can enhance accuracy).

\underline{Effect of weighting.} Figure~\ref{fig:a-vs-n-var-p2_utils} shows the effect of weighing (by utility values) in collaboration on the accuracy for different values of randomization noise, $p_2$. The devices with non-zero $p_2$ values are given a weight of 0.1 against the weight of 1 for $p_2=0$. From the figure we can observe that when $p_2$ is low and the number of devices with non-zero $p_2$ (other devices have $p_2=0$), $n_{noisy}$, is low, there is no much benefit from weighting as expected. However when $n_{noisy}$ is high, we can see that weighting gives more accuracy (about 4\%) than the base case (uniform weighting). Similarly when $p_2$ is high and $n_{noisy}$ is from 2 to 5, we can see that weighting gives advantage (about 5\% for $n_{noisy}=5$). However when $n_{noisy}=6$, there are too many noisy devices (around 86\%) (we assume this will not happen in reality) that weighting is unable to provide extra accuracy than the uniform case. We intentionally omitted the confidence intervals (which are within 2\%) in order to show the points clearly.

\underline{Accuracy by building type.} Figure~\ref{fig:a-vs-building-var-p2} shows the room-level and building-level accuracies for different types of buildings with different $p_2$ values (results for $p_1$ are similar). The lower accuracy for "Library" and "Student Center" cases is due to limited test data and possibly lack of additional feature variability. We note that these results are only to show the proof of feasibility and can be improved considerably with more amount of training and test data. 
We also mention that, apart from collaboration, \textit{coverage} of each provider (places the provider has visited) also affects accuracy of the solution. One can imagine that as number of places visited by the providers (i.e., database entries) increases, the accuracy of the framework increases. Even though we have this result we could not include due to lack of space. Finally, \textit{we have not compared our solution against other indoor localization solutions} for the same reason we mentioned in Sect.~\ref{sec:intro}---our solution neither competes nor is comparable with indoor localization solutions.

\balance

\section{Conclusion and Future Work}\label{sec:conc}
We presented a privacy-preserving, multi-modal, collaborative cross-building room-level localization framework.\jj We believe it is also a low-energy solution since we make use of Android system scans most of the time.\jj The proposed approach can achieve sub room level granularities though it is tested for room-level ($\approx 5m$) granularity. 
The results show that the proposed approach has the potential for wide-spread public usage (including integration into mobile OS) for privacy-preserving pervasive localization. 



As future work, we will perform the following: (1) increase the solution's current granularity by considering Geo-magnetic signal, and further exploiting the microphone of the device to both \textit{actively} and \textit{passively} probe the surrounding environment via sound signals (Channel State Information~(CSI) cannot be used as ours is mobile based solution); (2) evaluate requester $\leftrightarrow$ CTPM $\leftrightarrow$ providers communication along with ToR routing and secure MPC in a network simulator such as ns-3; (3) dynamically modify similarity threshold based on APs density.
\textbf{Acknowledgments:}
We thank the Department of Homeland Security~(DHS) Science \& Technology Directorate~(S\&T) Cyber Security Division for their support 
under contract No.~D15PC00159.

\balance

\bibliographystyle{IEEEtran}

\bibliography{references_v3.0,Mendeley}


\end{document}